\documentclass[a4paper,11pt]{article}

\usepackage{jheppub} 

\usepackage[T1]{fontenc} 

\usepackage{graphicx}
\usepackage{epsfig}
\usepackage{rotating}
\usepackage{amssymb}
\usepackage{subfigure}
\usepackage{dsfont}
\usepackage{psfrag}
\usepackage{amsmath,euscript,array,mathrsfs}
\usepackage{bbold}
\usepackage{epsf}
\usepackage{slashed}

\newcommand{\beq}{\begin{equation}}
\newcommand{\eeq}{\end{equation}}
\newcommand{\beqs}{\begin{eqnarray}}
\newcommand{\eeqs}{\end{eqnarray}}
\newcommand{\lsim}{\mathrel{\raisebox{-
.6ex}{$\stackrel{\textstyle<}{\sim}$}}}
\newcommand{\gsim}{\mathrel{\raisebox{-
.6ex}{$\stackrel{\textstyle>}{\sim}$}}}

\def\hbar{\hspace{0pt}\raisebox{1pt}{$-$} \hspace{-7pt} h}

\def\di{\mbox{d}}

\newcommand{\be}{\begin{equation}}
\newcommand{\ee}{\end{equation}}
\newcommand{\bea}{\begin{eqnarray}}
\newcommand{\eea}{\end{eqnarray}}

\def\lbldef#1#2{\expandafter\gdef\csname #1\endcsname {#2}}

\def\href#1#2{#2}

\newcommand{\ber}{\begin{eqnarray}}
\newcommand{\eer}{\end{eqnarray}}

\newcommand{\beqar}{\begin{eqnarray}}

\newcommand{\eeqar}{\end{eqnarray}}

\newcommand{\dsl}
  {\kern.06em\hbox{\raise.15ex\hbox{$/$}\kern-.56em\hbox{$\partial$}}}

\newcommand{\eeqarr}{\end{eqnarray}}
\newcommand{\ZZ}{{\rm \kern 0.275em Z \kern -0.92em Z}\;}

\def\CC{{\mathchoice
{\rm C\mkern-8mu\vrule height1.45ex depth-.05ex
width.05em\mkern9mu\kern-.05em}
{\rm C\mkern-8mu\vrule height1.45ex depth-.05ex
width.05em\mkern9mu\kern-.05em}
{\rm C\mkern-8mu\vrule height1ex depth-.07ex
width.035em\mkern9mu\kern-.035em}
{\rm C\mkern-8mu\vrule height.65ex depth-.1ex
width.025em\mkern8mu\kern-.025em}}}

\def\RR{{\rm I\kern-1.6pt {\rm R}}}

\def\ZZ{{\rm Z}\kern-3.8pt {\rm Z} \kern2pt}
\def\IB{\relax{\rm I\kern-.18em B}}
\def\ID{\relax{\rm I\kern-.18em D}}
\def\II{\relax{\rm I\kern-.18em I}}
\def\IP{\relax{\rm I\kern-.18em P}}

\newcommand{\bear}{\begin{eqnarray}}
\newcommand{\eear}{\end{eqnarray}}

\def\to{\rightarrow}

\def\to{\rightarrow}

\def\6{\partial}

\def\bea{\begin{eqnarray}}
\def\eea{\end{eqnarray}}

\def\beqx{\begin{displaymath}}
\def\eeqx{\end{displaymath}}

\newcommand{\bmat}{\left(\begin{array}}
\newcommand{\emat}{\end{array}\right)}

\def\bo{{\raise-.3ex\hbox{\large$\Box$}}}               
\def\face{{\raise.2ex\hbox{$\displaystyle \bigodot$}\mskip-2.2mu \llap {$\ddot
        \smile$}}}                                   
\def\>{\rangle}                                      
\def\<{\langle}                                      

\def\leftrightarrowfill{$\mathsurround=0pt \mathord\leftarrow \mkern-6mu
        \cleaders\hbox{$\mkern-2mu \mathord- \mkern-2mu$}\hfill
        \mkern-6mu \mathord\rightarrow$}        
\def\dvec#1{\vbox{\ialign{##\crcr
        \leftrightarrowfill\crcr\noalign{\kern-1pt\nointerlineskip}
        $\hfil\displaystyle{#1}\hfil$\crcr}}}           

\def\-{\hphantom{-}}

\title{Large mass hierarchies from strongly-coupled dynamics}

\author[a]{Andreas Athenodorou,}
\affiliation[a]{Department of Physics, University of Cyprus, POB 20537, 1678 Nicosia, Cyprus}

\author[b,c]{Ed Bennett,}
\affiliation[b]{Department of Physics, College of Science, Swansea University, Singleton Park, Swansea SA2 8PP, UK}
\affiliation[c]{Kobayashi-Maskawa Institute for the Origin of Particles and the Universe (KMI), Nagoya University, Nagoya 464-8602, Japan}

\author[d]{Georg Bergner,}
\affiliation[d]{Universit\"at Bern, Institut f\"ur Theoretische Physik,
Sidlerstr.~5, CH-3012 Bern, Switzerland}

\author[e]{Daniel~Elander,}
\affiliation[e]{National Institute for Theoretical Physics, School of
  Physics, and Mandelstam Institute for Theoretical Physics,
  University of the Witwatersrand, Johannesburg, Wits 2050, South
  Africa} 

\author[f,g]{C.-J. David Lin,}
\affiliation[f]{Institute of Physics, National Chiao-Tung University, Hsinchu
30010, Taiwan}
\affiliation[g]{CNRS, Aix Marseille Universit\'{e}, Universit\'{e} de Toulon,
Centre de Physique Th\'{e}orique,\\UMR 7332, F-13288 Marseille, France}

\author[b]{Biagio Lucini,}

\author[b]{ Maurizio Piai.}

\date{\today}

\abstract{
Besides the Higgs particle discovered in 2012, with mass 125 GeV, 
recent  LHC data show  tentative signals for new resonances in
diboson as well as diphoton searches at high center-of-mass energies (2 TeV and 750 GeV, respectively).
If these signals are confirmed (or other new resonances are discovered at the TeV scale),  the large hierarchies 
between masses of new bosons require a dynamical explanation.
Motivated by these tentative signals of new physics, 
we investigate the theoretical possibility 
that large hierarchies in the masses of glueballs could arise dynamically in new strongly-coupled
 gauge theories extending the standard model of particle physics.
We study lattice data on non-Abelian gauge theories in the
(near-)conformal regime as well as a simple toy model in the context of gauge/gravity dualities.
We focus our attention on the ratio $R$ between the mass of the lightest spin-2 and spin-0 resonances,
that for technical reasons is a particularly convenient and clean observable to study.
For models in which (non-perturbative) large
anomalous dimensions arise dynamically,
 we show indications that this mass ratio can be large, with
 $R>5$. { Moreover, our results suggest that $R$ might be related to
 universal properties of the IR fixed point. Our findings provide an
 interesting step towards understanding large mass ratios in the
 non-perturbative regime of quantum field theories with (near) IR conformal
 behaviour.}

}

\begin{document}
\maketitle
\flushbottom


\newpage
\section{Introduction}

The Higgs particle~\cite{Higgs} is the first example  in nature of 
a boson the mass of which is not protected by  symmetry arguments. Its low-energy Effective Field Theory (EFT)
description in terms of a weakly-coupled scalar field is fine-tuned, as 
additive renormalisation makes it sensitive to unknown physics up to high scales.
This is the hierarchy problem,
one of the main motivations to investigate theoretical extensions of the Standard Model (SM).

Searches at the LHC for resonant production of  particles decaying into diphoton or
diboson final states show  excesses
at center-of-mass energies around 750 GeV~\cite{diphoton} and 2 TeV~\cite{diboson}, respectively.
Such tentative signals  bring into further question the 
possibility of writing EFTs with hierarchical scales without invoking fine-tuning.

If a new strongly-coupled theory is responsible for electroweak
symmetry breaking and all the physical phenomena connected with it~\cite{TC,WTC}, 
it would provide an elegant and conclusive solution to the hierarchy 
problem(s) of the electroweak theory, in a context in which 
new composite states  appear at energies accessible to the LHC. 

A simple rescaled version of QCD cannot  explain 
the phenomenology we see at the electroweak scale and above.
Precision electroweak studies have already ruled out this possibility ---
a conclusion further supported by the discovery of the Higgs particle. 
A realistic model must exhibit dynamical properties
that are radically different from QCD, yielding
 large hierarchies in the masses of particles.

Non-perturbative methods are needed to test this broadly defined scenario.
Over the past decades, great progress has been made by using lattice gauge theories,
as well as gauge/gravity dualities.
The two approaches  
can be considered to be complementary to one another, as we will discuss later in the paper. 

We want to identify models that dynamically produce a large {\it mass hierarchy} 
between composite states, that cannot be explained in simple terms 
by symmetry arguments in a low energy EFT context (and without fine-tuning).
We are aiming at something more than what in QCD is captured by
 the chiral Lagrangian, or  Heavy Meson Chiral Perturbation Theory ($\chi$-PT), that explain the masses 
and properties of  pions and of heavy-light mesons, respectively. 
We want to find appropriate physical observables that allow such an identification to be
assessed in a clean, unambiguous way distinctive from simple arguments
formulated at weak coupling on the basis of internal symmetries. This
should be based on the recent developments in non-perturbative
techniques, both from lattice studies and in the context of
gauge/gravity duality. 

In this paper, we provide one interesting step in this direction, and suggest possible ways to further
develop this challenging research program in the future.
Our starting point is the observation that, irrespectively of the microscopic details,
all Lorentz-invariant four-dimensional field theories admit a stress-energy tensor $T_{\mu\nu}$.
Correlation functions involving $T_{\mu\nu}$ can be analysed in terms of their scalar (trace) part 
and tensor  (transverse and traceless) part.
A  well-defined observable  is  the ratio
\beqs
R&\equiv&\frac{M_T}{M_0}\,,
\eeqs
where $M_T$ is the mass of the lightest spin-2 composite state, while $M_0$ is the mass of the lightest spin-0 state.
This quantity is defined universally, it can be computed explicitly in a wide variety of models, 
it is scheme-independent, and it is not directly controlled by  internal global symmetries
of the theory. It is legitimate to compare $R$ computed in theories with completely different
internal symmetries and symmetry-breaking patterns. This is a particularly welcome feature in the context of 
gauge theories with fermionic field content, where the physics of chiral symmetry and its breaking
introduces non-trivial model-dependent features.

In recent years a large number of different theories with non-QCD-like dynamics
 have been studied on the lattice (see the reviews~\cite{reviewlattice}). Much emphasis has gone into 
 the discussion of their mesonic properties, while technical difficulties have so far hindered the progress
 in understanding the glueballs of the theories, with the exception of some particularly neat cases.
Among the latter,  pure Yang-Mills $SU(N)$ theories are comparatively well understood, and the spectra 
of glueballs of various quantum numbers are known. The ratio $R$ has been computed  to be $1.4\lsim R_{\mathrm{YM}}\lsim 1.7$ 
with the lower bound reached for small $N$ and the upper bound by extrapolating to large $N$~\cite{glueballs,Lucini:2001ej,Chen:2005mg}. 
The question we want to address is whether there exist models in which $R\gg R_{\mathrm{YM}}$. 

We focus on $SU(2)$ theories with adjoint matter,
for which there are indications that the dynamics is IR conformal~\cite{DelDebbio:2009fd,DelDebbio:2010hu,DelDebbio:2010hx,Bursa:2011ru,DelDebbio:2015byq,SU2spectrum,SU2running,Lucini:2009an,SU2anomdim,Patella:2012da,Lucini:2013wsa,Athenodorou:2014eua}.
As explained below, the investigation of a conformal theory with
numerical simulations\footnote{Needless to say, on a finite lattice and at finite
  fermion mass, the spectrum will be different from that of the
  continuum massless theory. In this paper we collectively call {\it lattice artefacts}  all  effects
  that are not present in the continuum massless field theory,
  inclusive of finite volume, finite spacing and finite fermion mass
  effects.} requires an extrapolation to the zero residual mass,
infinite volume, and continuum limit. In this conformal limit all
masses are zero. It is, nevertheless, possible to extrapolate a finite
value of $R$ in an unambiguous way. We will show, also by comparison
with our toy model, that this procedure depends on the precise way the
different limits are taken. We are also interested to see whether the
ratio $R$ obtained in this way contains non-trivial information about 
the anomalous dimensions,  generalising~\cite{DelDebbio:2010hx}. 

{
By exploiting arguments originating from  scale invariance, the authors of~\cite{DelDebbio:2010hx}
derived scaling relations for spectral masses in a mass-deformed IR conformal gauge
theory as a function of the mass deformation $m$. 
In particular, all spectral scales 
follow the same power law in $m$ when approaching the massless limit, with the exponent being
determined solely by the anomalous dimension of the deforming operator. 
In general, the coefficients in front of such a power law depend on the state considered and on details of the theory.
However, scale invariance in the IR might provide extra 
constraints on them. 

Consequences of scale invariance are well studied
in statistical systems. One of the main results is universality, which
states that at a second order phase transition (where the theory
is scale invariant) certain properties of a Statistical Mechanics
system, including the critical exponents (anomalous dimensions), 
depend only on the symmetries of the Hamiltonian and on the
dimensionality, and not on microscopic details such as the specific elementary
degrees of freedom.  

Universality applies also to less familiar examples. 
The power law approach (as a
function of deformation parameters such as 
temperature or external magnetic field) to the
critical point is governed by model-dependent coefficients.
But there are particular combinations that only depend on symmetries and
dimensionality, and are hence  universal. For a detailed discussion
pertaining to the example of the two-dimensional Ising model, we refer
the reader to~\cite{Delfino:2003yr}. 

Here, we would like to stress
that, while explicit calculations can be carried out mostly in two
dimensions, the concept of {\em universal
  amplitude ratios} is more general and descends from the concept of
universality (see e.g.~\cite{Cardy:1996xt}). It is hence a legitimate
question to ask whether there are universal  ratios in IR
conformal gauge theories, and in particular whether $R$ (which, we
stress, can be defined in many local QFTs) is one of them.
} 

Gauge/gravity dualities allow the use of weakly-coupled classical field theory coupled to gravity 
in higher dimensions to explore the dynamics of strongly-coupled 
four-dimensional theories~\cite{AdSCFT,reviewAdSCFT}. 
There exists a fully algorithmic procedure for computing glueball spectra,
at leading order in the large-$N$ and large 't Hooft coupling limits.
It uses the action of a sigma-model coupled to gravity in five dimensions,
that in the top-down approach may be obtained as a consistent truncation of the dimensional reduction of a
more fundamental gravity theory in higher dimensions~\cite{BHM,E,EP}.

These techniques have been applied successfully to the dual of confining gauge theories,
such as the Witten model~\cite{W,BMT,EFHMP}, the Klebanov-Strassler model~\cite{KS,BHM}, the Maldacena-Nunez 
model~\cite{CVMN,BHM} and several generalisations, including  cases in which
 the mass spectra include an 
 anomalously light state~\cite{NPP,ENP,E2014}.
For instance,  the Witten model has many properties that make it resemble the gravity dual of a 
non-supersymmetric Yang-Mills theory, and the gravity calculation yields $R\simeq 1.7$~\cite{BMT}.

Less sophisticated models exist for which the asymptotic high-energy behaviour of the theory is simpler to interpret,
but in which the geometry is not smooth, indicating that these models provide  incomplete
descriptions of  long-distance physics. An example is the GPPZ model~\cite{GPPZ} (see also~\cite{PW}),
in which the dual field theory is ${\cal N}=4$ super-Yang-Mills, deformed by a particular
symmetry-breaking mass term. The (gravity) calculation of the glueball spectrum yields $R=\sqrt{2}\simeq 1.4$~\cite{MP}.

We are interested in modelling with gravity a physical situation that is similar to the 
one found on the lattice: a CFT is deformed by a relevant coupling, thus
inducing a departure from AdS of the gravity background, 
ultimately leading to the geometry ending along the extra dimension. 
The spectrum consists of masses that all scale in a universal way  with the deformation scale~\cite{DelDebbio:2010hx},
and hence $R$ can be given a physical meaning.
While the individual coefficients controlling the masses are model-dependent, we want to investigate 
whether the ratio $R$ shows universal properties.
In the absence of a known class of supergravity backgrounds that 
describe the deformation of a CFT with  tuneable  anomalous dimension,
we resort to a toy model, in the spirit of
the bottom-up approach to holography.

{
The main idea of this paper is the following.
We consider a set of conformal theories that admit a relevant deformation,
and compute the spectra of scalar and tensor bound states in the presence of this
deformation. 
Inspired by the idea of universality, in particular of universal amplitude ratios borrowed from statistical field theory,
we question whether it is possible that the ratio $R$  defined earlier might be a quantity exhibiting 
such universal properties.

We compare the results obtained from lattice $SU(2)$ theories  with two different 
field contents, and hence two different anomalous dimensions $\gamma^{\ast} \equiv
\Delta - 1$ for the $\bar{\psi}\psi$ operator,
to results obtained in a completely different class of models, built within the bottom-up approach to holography.
In the latter class of theories, $\Delta$ is a tuneable parameter: we compute the value of $R$ for generic $\Delta$,
and compare to the results of the $SU(2)$ lattice theories for which the same values of $\Delta$ are available.
As we will see, we find a surprisingly good level of agreement, though subject to numerous caveats that we discuss 
in detail.

Only with further future work on the subject it will be possible
to ascertain whether this is an indication of a strong form of universality, 
manifesting itself in the fact that $R$ is just a function of $\Delta$
(and the space-time dimensionality),
but not of the details of the theory, or whether the agreement we uncover is a more modest
result of generic similarities specific to the theories we analysed.
We will comment on further steps we suggest for future work in this direction at the end of the paper.
}

The paper is organised as follows.
In Section~\ref{Sec:lattice} we describe the lattice results for $R$ in two 
models, both based on $SU(2)$ gauge group, in which the field content consists of either one or two 
Dirac fermions transforming in the adjoint representation.
In Sec.~\ref{Sec:string} we illustrate the results for $R$ as a function of the dimension of the operator 
deforming a CFT for a toy model built as a generalisation of the five-dimensional consistent truncation
leading to the GPPZ background.
In Sec.~\ref{Sec:physics} we analyse the physical implications of our results, and extract from them
some useful lessons of general relevance.
We critically discuss the limitations intrinsic in our work 
and  we outline future avenues for research that might overcome these limitations in Sec.~\ref{Sec:outlook}.

\section{Lattice $SU(2)$ gauge theories with adjoint matter}
\label{Sec:lattice}
\subsection{Lattice formulation and setup}
\label{Sec:lattice:formulation}

The $SU(2)$ gauge theory with two Dirac fermions in the adjoint
representation of the gauge group is the first that has been shown to be infrared (IR) conformal
in the context of lattice studies of the conformal window.
A vast body of literature exists on its
spectrum~\cite{DelDebbio:2009fd,DelDebbio:2010hu,DelDebbio:2010hx,Bursa:2011ru,DelDebbio:2015byq,SU2spectrum},
on the running of its coupling~\cite{SU2running} and on the anomalous dimension 
of its chiral condensate~\cite{Lucini:2009an,SU2anomdim,Patella:2012da}. 
From a phenomenological point of view, this theory is likely to be of limited
relevance, since the anomalous dimension of the condensate is small.
Yet the study of this
gauge theory has been a crucial milestone in numerical explorations
of strongly interacting dynamics beyond the Standard Model, 
fostering the development of  specific investigation techniques for nearly
conformal gauge theories discretised on a spacetime lattice. 

It has been
shown in~\cite{Athenodorou:2014eua} that the $SU(2)$ gauge theory with a single
adjoint Dirac flavour (or, equivalently, two Majorana flavours) is near the onset of the conformal window and
has an anomalous dimension of order one. This makes it  an
ideal lattice playground for non-perturbative tests of near-conformal gauge
theories with  large anomalous dimensions.

In Minkowski space, the Lagrangian of an $SU(2)$ gauge theory coupled to
$N_f$ flavours of adjoint Dirac fermions of mass $m$ is 
\begin{eqnarray}
\label{eq:actcont}
{\cal L} = \sum_{i=1}^{N_f} \overline{\psi}_i \left( i \slashed{D} - m \right) \psi_i
- \frac{1}{2} \mathrm{Tr} \left[ G_{\mu \nu} G^{\mu \nu} \right]\ ,
\end{eqnarray}
where $\slashed{D} \equiv \left(\partial_{\mu} + i g A_{\mu} \right)
\gamma^{\mu}$, $\gamma^{\mu}$ are the Dirac matrices, $A_{\mu} \equiv
\sum_a T^a A_{\mu}^{a}$ with $a = 1,2,3$, and $T^a$ are the $3\times 3$
generators of $SU(2)$ in the adjoint representation. The field strength tensor
is $G_{\mu \nu} = \partial _{\mu} A_{\nu} - \partial
_{\nu} A_{\mu}+ i g [A_{\mu}, A_{\nu}]$, with $g$ the
coupling. The trace is 
over the gauge indexes, with the generators $T^a$ normalised such
that $\mathrm{Tr}(T^a T^b) = \delta_{ab}/2$. We shall consider the
cases $N_f = 1$ and $N_f = 2$.

The action on the 
Euclidean spacetime grid is
\begin{eqnarray}
S=S_{\mathrm{G}} + S_{\mathrm{F}}\,,
\end{eqnarray}
where
\begin{eqnarray}
  S_{\mathrm{G}} = \beta \sum_{p} {\rm Tr} \left[ 1 - U(p)  \right] \ ,
\end{eqnarray}
with $U(p)$ the lattice plaquette, is the pure gauge part (referred to as the Wilson plaquette action),  and
\begin{eqnarray}
S_{\mathrm{F}} =  \sum_{x,y} \sum_{i=1}^{N_f} {\overline \psi}_i (x) D(x,y)\psi_i (y)
\end{eqnarray}
is the fermionic contribution. The massive Dirac operator $D(x,y)$ in the
Wilson fermion discretisation used throughout this work is
\begin{align}
  D(x,y) = \delta_{x,y} 
  - \kappa &\left[ \left(1-\gamma_{\mu}  \right) U_{\mu} (x) \delta_{y,x+\mu} 
    \left(1+\gamma_{\mu}  \right) U^{\dagger}_{\mu} (x-\mu)
      \delta_{y,x-\mu}   \right] \ ,
\end{align}
with $\gamma^{\mu}$ the Euclidean Dirac matrices. $x$ and $y$ are points on
the lattice. In the previous equation, the lattice links $U_{\mu}(x)$
are written in the adjoint representation. $\kappa = 1/(8+2 a m)$ is the hopping parameter,
with $a$ the lattice spacing and $m$ the bare fermion mass.  For computational
reasons, in numerical simulations one has to consider the theory in the presence of a non-zero
mass $m$. The behaviour of the model as the mass is taken to zero allows
to disentangle between the near-conformal and the chiral-symmetry
broken cases. 

The path integral of the theory is 
\begin{equation}
Z = \int {\cal D} U {\cal D} \overline{\psi} {\cal D} \psi
e^{-S} \ , 
\end{equation}
and the vacuum expectation value of any operator $O(U,  \psi, \overline{\psi})$ is given by
\begin{equation}
\langle O \rangle  = \frac{1}{Z} \int  {\cal D} U {\cal D}
  \overline{\psi} {\cal D} \psi O(U,  \psi, \overline{\psi}) e^{-S} \ .
\end{equation}
If $O$ carries specific $J^{PC}$ quantum numbers,
at large distance $r = |x - y| $ the correlator
\begin{equation}
{\cal C}(r) = \langle O^{\dag} (x) O(y) \rangle \ , 
\end{equation}
decays as
\begin{equation}
\label{eq:correlator}
{\cal C}(r) = k_1 e^{- M_1 r} \,+ \,k_2 e^{-M_2 r} \,+\,\cdots \ ,
\end{equation}
with $M_i$ the lowest spectral masses in the given $J^{PC}$ channel. 
For large enough $r$, only the contribution from the state with lowest
mass $M_1$ survives, provided the coefficient $k_1$ 
(related to the decay constant) is not suppressed. {Note that in the
previous expression the invariant mass $M_1$ is not associated
necessarily to a stable state or a resonance, but could correspond to a
scattering state. For instance, in QCD, for which the mass of the
vector is above the threshold of the two pion system, unless the
corresponding weighting coefficient $k_1$ is suppressed for dynamical
reasons or by an appropriate choice of the probe operators used
in the calculation, in general with this technique one would extract
the invariant mass associated to the scattering of two pions with $J =
1$, and the mass of the resonance will only manifest as an excitation~\cite{Maiani:1990ca}.
The latter discussion about this somewhat technical aspect of the calculation
will be  relevant when analysing the numerical results obtained for
the investigated lattice theories.}

We shall focus on glueball masses,\footnote{Due to the choice of the probe, in
  the following we will refer to the corresponding spectral states as
  glueballs, although one has to keep in mind that these states will
  mix with meson-like states with the same quantum numbers.} that following
consolidated procedures are extracted using a
variational calculation including spatial Wilson loops of various
sizes and shapes transforming in the irreducible representation of the 
symmetry group of the cube. Continuum spins are then reconstructed by
looking at the embedding of this group into the continuum rotation
group. Further technical details on how glueball masses are extracted
using this method are provided for instance in~\cite{glueballs}.  

\begin{figure}
\includegraphics[width=0.95\columnwidth]{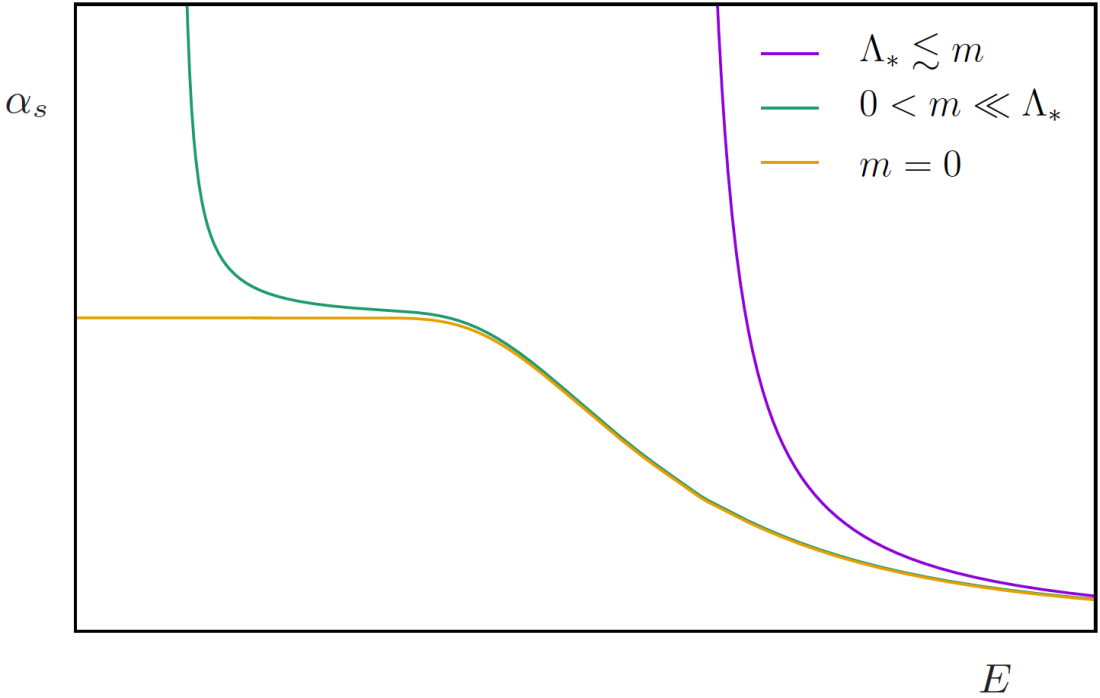}
	\caption{Schematic representation of $\alpha_s  = g^2/(4\pi)$ as a function of $E$ for  three representative choices of fermion mass $m$. 
	In yellow the massless $m=0$ case, in which the only scale is $\Lambda_{\ast}$. 
          In purple a case in which the mass  $m$ is very large.
           In green a case in which $m\ll \Lambda_{\ast}$, and hence
           $\Lambda_0\ll \Lambda_{\ast}$. \label{fig:deformed}}
\end{figure}

\subsection{Mass deformation and infrared behaviour}
\label{sect:lattice:masdef}

The claimed IR-conformality of the two theories of interest is true only in the chiral limit.
In the presence of a mass deformation, we must understand how the would-be
IR-conformal theory reacts to the explicit breaking of scale
invariance introduced via $m\neq 0$.\footnote{With abuse of notation, we
  denote with $m$ both the Lagrangian bare mass and the
  renormalised fermion mass, with the latter being the one that is
  relevant for the present discussion. A prescription to introduce the
  renormalised mass is given in~\cite{Miransky:1998dh,DelDebbio:2010hx}.} 

We show in Fig.~\ref{fig:deformed} a cartoon of the running coupling $\alpha_s\equiv \frac{g^2}{4\pi}$ as a function of the energy scale $E$,
for various choices of $m$, in order to illustrate the various possible regimes of $m$ that
yield physically distinctive features.
Firstly, in the $m=0$ case the IR-conformal behaviour  features a dynamical scale
$\Lambda_{\ast}$ that separates the perturbative regime from
the long-distance behaviour. The latter is characterised by the asymptotic approach
of the coupling to a finite value $g^{\ast}$ at low $E$. 
The scale $\Lambda_{\ast}$ may be estimated on the basis of perturbation theory,
and is  the lowest scale $E$ at which the dynamics is well
captured by the physics of the trivial UV fixed point.

If we deform the theory with a large mass $m$, chosen to be much larger than the  dynamical scale $\Lambda_{\ast}$ 
of the massless theory, 
we  completely destroy infrared
conformality, giving rise to a confining behaviour similar to that
of a Yang-Mills theory with heavy 
quarks. The scale at which the theory confines is $\Lambda_0 \neq \Lambda_{\ast}$, 
and at this scale the fermions are effectively
decoupled.

For small mass $m\ll \Lambda_{\ast}$, the breaking term acts as a soft deformation.
The theory behaves as IR-conformal down to some energy of the
order of a scale $\Lambda_0$, which is determined in a
dynamically non-trivial way by the mass deformation $m$. For $E\sim \Lambda_0$ 
the confining behaviour reappears. An explicit calculation  is possible
when the fixed point is weakly-coupled~\cite{Miransky:1998dh}. In this
case, one finds  $\Lambda_{0} \simeq m e^{- \frac{1}{b_0 \alpha_s^{\ast}}}$
with $b_0 = (11/6 \pi) N_c$ ($N_c$ being the number of colours) and
$\alpha_s^{\ast} = g^{\ast 2}/(4 \pi)$ the value of the coupling at the fixed point. 
The mass spectrum is determined by
$\Lambda_{0} $ and the meson masses are enhanced by a factor $e^{\frac{1}{b_0 \alpha_s^{\ast}}}$~\cite{Miransky:1998dh,DelDebbio:2009fd}. 

Regardless of the strength of 
$g^{\ast}$, in the energy region  
\begin{equation}
\Lambda_{0} \ll E \ll \Lambda_{\ast}\ ,
\end{equation}
in which the mass-deformed theory behaves as if it were IR conformal,
all spectral masses $M_i$
scale as
\begin{eqnarray}
M_i \propto m^{\frac{1}{\Delta}} \ ,
\end{eqnarray}
where $\Delta =1 + \gamma^{\ast}$ and $ \gamma^{\ast}$ is the
anomalous dimension of the chiral condensate~\cite{DelDebbio:2010jy}.  

When the theory is formulated on a spacetime lattice, an ultraviolet
cutoff $a\equiv\frac{1}{\Lambda_{\mathrm{UV}}}$ is introduced. 
The value of $a$ can be tuned by changing
the coupling $\beta$. 
The continuum
limit itself is realised at $\beta = \infty$. Computing
observables numerically causes the introduction of an infrared cutoff
related to the lattice size $L = n a\equiv \frac{1}{\Lambda_{\mathrm{IR}}}$, with $n$ the
number of lattice points in the considered direction. 
Physical
observables must  not be
affected by cutoff effects. In the case of a particle of mass $M$,
this  requires
\begin{equation}
\label{eq:artefacts}
a \ll \frac{1}{M} \ll L \,.
\end{equation}
 Conversely, if $1/M \lsim a$ the state is affected by discretisation
artefacts, while if $1/M \gsim L$ finite volume effects  dominate the
calculation.

When analysing the approach of lattice data to the chiral
point, in addition to the mass deformation, the infrared cutoff
scale $L$ needs to be taken explicitly into account. Borrowing a
consolidated analysis technique from the investigation of critical
phenomena, the finite size can be seen as a
renormalisation group relevant direction with mass dimension $-1$. The
dimensionless variable describing scaling with $m$ and $L$ is then $x =
L m^{1/\Delta}$. From the Widom form of the effective action, one can
derive the scaling
law~\cite{Lucini:2009an,DelDebbio:2010hu,DelDebbio:2010hx,DelDebbio:2010ze}  
\begin{equation}
\label{eq:scaling}
L M_i = f_i (x)\ , 
\end{equation}
i.e.\ spectral masses in units of the lattice size are universal
functions of the scaling variable. One can use the lowest-order
relation for a particle of mass $M_0$ 
\begin{equation}
L M_0 \propto x 
\end{equation}
to rewrite the scaling in Eq.~(\ref{eq:scaling}) as~\cite{Lucini:2009an}
\begin{equation}
L M_i = f_i (L M_0) \ , 
\end{equation}
with ratios of spectral masses assuming a universal form:
\begin{equation}
M_i/M_j = f_i (L M_0)/f_j(L M_0) \ . 
\end{equation}
This scaling relation is particularly useful for an unbiased and direct comparison
of lattice data obtained at different values of $m$ and $L$, as it
accounts for the finite size without making explicit reference
to the anomalous dimension. At large $L$, the leading
behaviour with $M_0$ of the previous equation predicts constant mass
ratios as $m$ is varied~\cite{DelDebbio:2009fd,DelDebbio:2010hu,DelDebbio:2010hx}.

Equipped with these considerations, in
subsections~\ref{Sec:lattice:twoflavours}~and~\ref{Sec:lattice:oneflavour}
we present lattice results for the ratio $R \equiv M_T/M_0$
with $M_T$ and $M_ {0}$ respectively
the mass of the tensor and scalar, determined by probing the theory with
glueball-like operators.

\subsection{The femto-universe}
\label{Sec:femto}

Since we will comment also on predictions 
in the small-volume limit, it is convenient to discuss some
generic features of the glueball spectrum obtained in this limit for Yang-Mills theories.  
Here we are referring to the scenario of the {\it
  femto-universe}~\cite{Bjorken:femto}, which is realised when the
size of the system is smaller than the shortest intrinsic (dynamically generated) 
length scale in the theory~\cite{L}.  In the case of QCD, this implies
\begin{equation}
\label{eq:femto_qcd}
 1/L > \Lambda_{{\rm QCD}} . 
\end{equation}
In this limit, the spectrum of an asymptotically-free gauge theory can
be extracted with perturbation theory.  

In this section, we restrict our discussion to the case of a
(3+1)-dimensional hypercubic volume $L^{3} \times T$, with $L$
satisfying the corresponding condition in Eq.~(\ref{eq:femto_qcd}).
Perturbation theory in such small cubic boxes can be complicated by
the global toroidal structure (``torons'') of the periodic
lattice~\cite{GonzalezArroyo:1981vw,Coste:1985mn}.  This complication
can be removed by employing the colour-twisted boundary
conditions (TBC)~\cite{'tHooft:1979uj,Coste:1986cb}.  To keep our discussion
simple, we  concentrate on results obtained using
TBC in Refs.~\cite{GonzalezArroyo:1988dz,Daniel:1989kj,Daniel:1990iz}.

The glueball spectrum of the pure Yang-Mills theories in the femto-universe with TBC takes the generic
form 
\begin{equation}
\label{eq:femto_glueball_generic}
 M_{G} = \frac{X_{0}}{L} + \frac{g^{2} X_{1}}{L} + \ldots , 
\end{equation}
where $g$ is the coupling, and $X_{0}$ is a constant that is completely determined by the
geometry and the boundary condition of the finite volume, as well as the cubic-group representation of
the glueball state.   Eq.~(\ref{eq:femto_glueball_generic}) is
the result of perturbation theory.  
With a specific choice of the twist that
preserves the cubic symmetry, $X_{0}$ takes the same value for all the
states associated with the irreducible representations of the cubic
group~\cite{Daniel:1989kj}.   In particular, this means that the
scalar and the tensor glueballs are degenerate, $R=M_{T}/M_{0} = 1$,  in the $g^{2}
\rightarrow 0$ limit.  We stress that this value of the mass ratio, $R$, is the consequence of
the box geometry, the boundary condition and group theory.  It does not result from the
underlying dynamics.

The gauge-field dynamics begins to set in at the first non-trivial
order in the expansion of Eq.~(\ref{eq:femto_glueball_generic}).  In
general, the coefficient $X_{1}$ depends on the spin of the
glueball.  For the case of TBC considered in Ref.~\cite{Daniel:1989kj},
it increases $R$.\footnote{This statement is also found to be valid
  when lattice artefacts are accounted for in the perturbative
  calculations~\cite{Daniel:1990iz}.}   
This coefficient depends on
the boundary conditions as well.  In fact, in the computation employing
periodic boundary conditions, the authors of
Refs.~\cite{Luscher:1982ma,F} find that the tensor glueball is lighter
than the scalar state at the first non-trivial order of perturbation
theory, although the mass ratio, $R$, is also close to unity.

In the femto-universe, the light-fermion masses are well below the
scale $1/L$.  Therefore, practically they can be considered as
massless.  Since the fermions couple to the pure gauge
degrees of freedom only perturbatively, they will not have any
significant (non-perturbative) effects on the glueball spectrum
discussed above.  The authors of Ref.~\cite{Daniel:1990iz} computed
the fermionic contribution in continuum perturbation theory to
$O(g^{2})$, and found it to be small.

\subsection{Numerical results for the two flavour theory}
\label{Sec:lattice:twoflavours}

\begin{figure}[t]
\includegraphics[width=0.95\columnwidth]{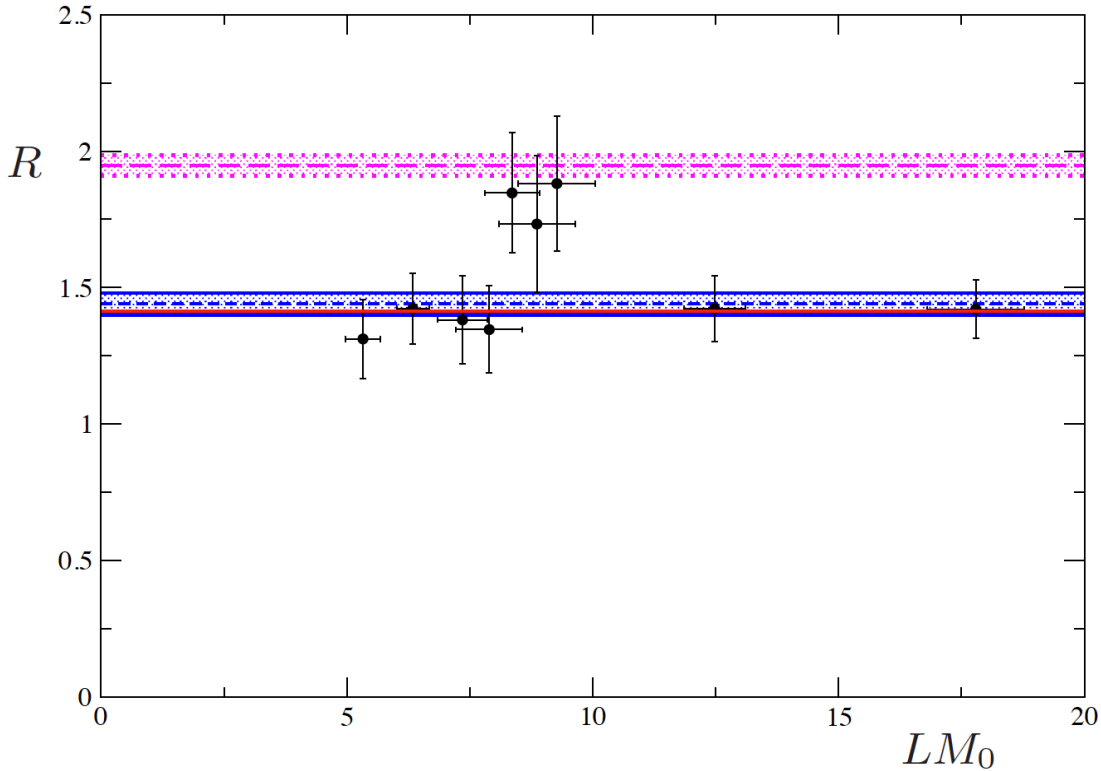}
	\caption{The ratio $R$ as a
          function of $L M_0$ for the $SU(2)$ gauge theory with two
          adjoint Dirac fermions. The horizontal blue band indicates the
         $SU(2)$  Yang-Mills value $R = 1.44(4)$. The pink band $R=1.95(4)$,
         and the red line at $R=\sqrt{2}$
         are discussed in
         Sec.~\ref{Sec:string}. The interpretation of the
         data is explained and discussed in Section~\ref{Sec:lattice:twoflavours}. \label{fig:fitrvsm0PP_2}}
\end{figure}

The construction of the interpolating operators for the measurements of the bound state spectrum
in adjoint QCD is similar to QCD with fermions in the fundamental representation. The physics of these
bound states is, however, much different. One important difference is the fact that, due to the real
representation, the massless theory has $SU(2 N_f)$ chiral symmetry that is broken to $SO(2N_f)$ by the fermion condensate.

The spectrum of the
theory has been widely studied, with an accurate analysis of possible
systematic effects related to the choice of interpolating operators and
to finite size effects given in~\cite{Bursa:2011ru} and a careful large volume
extrapolation presented  in~\cite{DelDebbio:2015byq}. { It is worth
noting that $M_0$ is the lightest state of the
theory and $M_T < 2 M_0$. Hence, the mass extracted in the tensor channel
is associated with a stable particle.} For small
fermion masses, the spectrum has spectral mass ratios that are
constant as a function of $m$. This signals (near-)conformal 
behaviour. Studies of the running of the coupling have exposed the
presence of an infrared fixed point. The anomalous dimension of the
condensate has been found to be $\gamma^{\ast} =
0.371(20)$~\cite{Patella:2012da,DelDebbio:2015byq} (see also~\cite{Miransky:1998dh,DelDebbio:2009fd,DelDebbio:2010hx}). 

The most accurate studies for the mass spectrum have been done at only
one value of the lattice coupling $\beta = 2.25$.
Taking the data from the simulations available in the
literature (mostly~\cite{DelDebbio:2009fd,DelDebbio:2010hx}
supplemented with the two largest volumes at $a m = -1.05$ given
in~\cite{DelDebbio:2015byq}),  we plot the ratio $R$  as a
function of $L M_0$ in Fig.~\ref{fig:fitrvsm0PP_2}. We show also the pure $SU(2)$ Yang-Mills value
$R \simeq 1.44(4)$ as determined by lattice calculations~\cite{Lucini:2001ej}  with a
horizontal line. A manifest feature of the data is the appearance of three distinct plateaus in $R$.
For the largest values of $L M_0$ we find good agreement with the Yang-Mills value.  
At intermediate values of $L M_0$ there is evidence of an enhancement of $R>1.4$.
At the smallest available values of $L M_0$, $R$ is systematically below the pure Yang-Mills value,
although still compatible with $R=1.44(4)$ within the large errors.

We conclude this subsection by noting that for the smallest volumes
at $a m = -1.15$ studied in~\cite{DelDebbio:2015byq} $R$ is compatible with the
expected femto-universe results. The explicit observation of the
deviation from the femto-universe regime in the results discussed here
is a good indication of absence of severe small volume effects for the
data presented in Fig.~\ref{fig:fitrvsm0PP_2}.

\begin{figure}[t]
	\includegraphics[width=0.95\columnwidth]{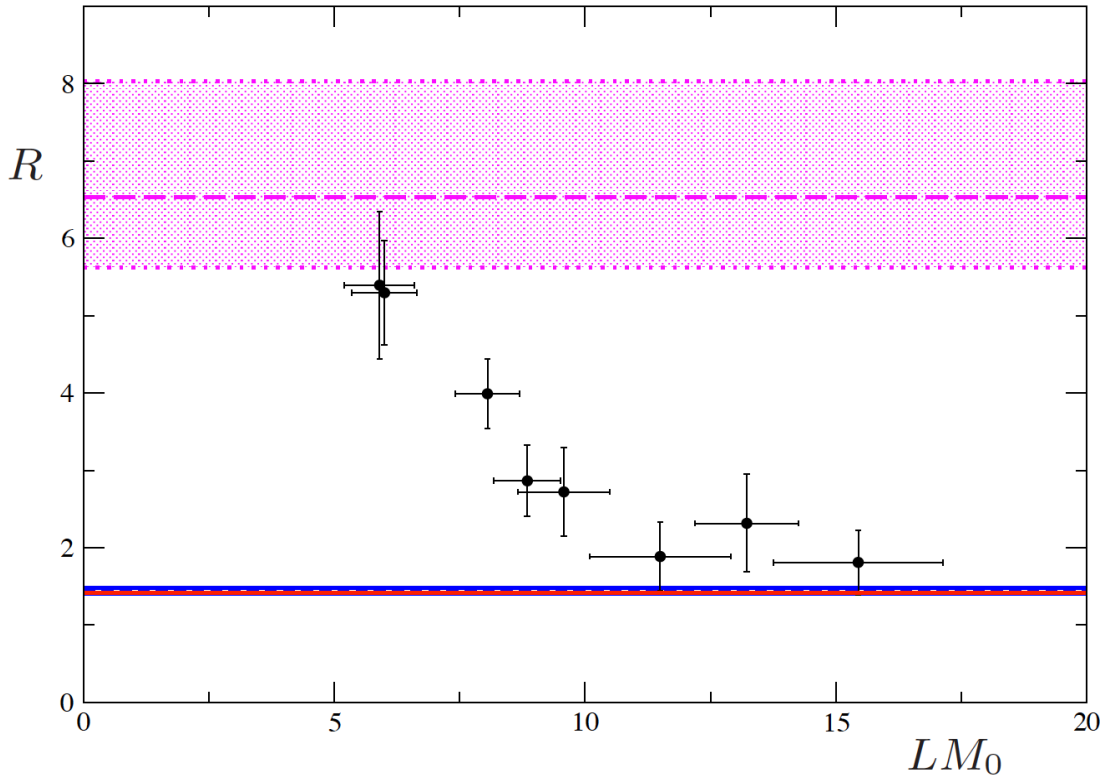}
	\caption{The ratio $R$ as a
          function of $L M_0$ for $SU(2)$ gauge theory with one
          adjoint Dirac fermion. The horizontal blue band indicates the
          SU(2) Yang-Mills value $R = 1.44(4)$. { The pink band at $R=6.53^{+1.50}_{-0.91}$}
          and the red line at $R=\sqrt{2}$
         are discussed in Sec.~\ref{Sec:string}. The interpretation of the
         data is explained and discussed in Section~\ref{Sec:lattice:oneflavour}. \label{fig:fitrvsm0PP_1}}
\end{figure}

\subsection{Numerical results for the one flavour theory}
\label{Sec:lattice:oneflavour}

In the $N_f = 1$ adjoint flavour theory, 
upon decomposition of the Dirac fermion into its two
Majorana constituents, the $SU(2)$ chiral symmetry
is manifest in the action. It is broken to $SO(2)$ if a non-zero
fermion condensate forms. Hence, chiral symmetry breaking would result
in the production of two Goldstone bosons.
A detailed analysis of the symmetries and
numerical results for various spectral states are
provided in~\cite{Athenodorou:2014eua}. In particular two independent degenerate
signals for the scalar channel, the
$0^{++}$ glueball and the mesonic isoscalar, have been considered.

The simulations (performed for the single value of  $\beta = 2.05$)
show again a spectrum with a markedly different signature than one would
expect in the chiral symmetry broken case. Ratios of
the lowest-lying spectral masses appear to be constant as a function
of the mass deformation $m$, with the lightest
state being a $0^{++}$ scalar. The would-be Goldstone bosons 
expected from the anticipated symmetry breaking pattern $SU(2)
\to SO(2)$ look like other ordinary massive states. The
condensate anomalous dimension has been found to be $\gamma^{\ast} =
0.925(25)$.   

We report in Fig.~\ref{fig:fitrvsm0PP_1} numerical
results for the ratio $R$ as a function of $L M_0 $, 
using the data for the $0^{++}$ glueball obtained
in~\cite{Athenodorou:2014eua} supplemented by additional calculations
that will be discussed elsewhere. While at the largest available values of $L M_0$
the ratio $R$ is compatible with the pure $SU(2)$ Yang-Mills theory, 
the ratio is significantly different from it for  lower $M_0 L$.  In
this latter regime, $R$ is considerably enhanced, even  with respect to  the
peak value for the $N_f = 2$ case.  

{
In the light of the discussion of potential contributions of scattering
states to the two-point function (see Section~\ref{Sec:lattice:formulation}), the existence of a region where $M_T >
2 M_0$ deserves further comments. In fact, one should expect that if
the resonance in the tensor channel has a mass that is much larger than that
of the scalar, the tensor correlator will be asymptotically
dominated by scattering states~\cite{Maiani:1990ca}. Hence, one might wonder whether we
have identified the wanted resonance or we are observing some spurious
object. 

In order to provide a convincing answer to this question, 
calculations with extended statistics and purpose designed analysis methods~\cite{Luscher:1985dn} 
are needed, which would go well beyond our current aims. Our
preliminary investigation, based on the scaling of the mass with the
lattice volume and on the expected degeneracy of the continuum tensor state in
two representations of the rotational group of the cube in the case of
a single particle, suggests that
the extracted mass identifies a resonance. 
The apparent absence of
scattering state contributions in the correlator might be due to the particular
construction of the trial variational operators, which are
optimised for single particles. 
}

\section{A string-inspired toy model and the dual mass spectrum}
\label{Sec:string}

We want to model  the dynamics of a conformal gauge theory in four dimensions,
in which the insertion of a relevant deformation via the coupling of an operator ${\cal O}$ of dimension $4-\Delta$
introduces a scale in the theory that discretises the spectrum and introduces a mass gap.
In the spirit of bottom-up holography, we consider a five-dimensional
sigma-model  consisting of one scalar $\Phi$ coupled to gravity.
The 5-dimensional action is
\beqs
\int\di^4x\di r\sqrt{-g}\left[\frac{R_5}{4}-\frac{1}{2}g^{MN}\partial_M\Phi\partial_N\Phi-V(\Phi)\right]\,,
\eeqs
where $R_5$ is the Ricci scalar in five dimensions, $V(\Phi)$ is the scalar potential
and $g_{MN}$ the five-dimensional metric.
The dynamics of the (canonically normalised) bulk scalar field descends from a superpotential for which,
 taking inspiration from  GPPZ~\cite{GPPZ,PW},
we reconsider the toy model proposed in~\cite{EP}:
\beqs
W&=&-\frac{3}{4}\left(1+\cosh 2\sqrt{\frac{\Delta}{3}}\Phi\right)\,,
\eeqs
where $\Delta$ is the scaling dimension of the parameter controlling the deformation. The scalar potential is
\beqs
V&=&\frac{1}{2}\left( W_{\Phi}\right)^2\,-\frac{4}{3}W^2\,,
\eeqs
where $W_{\Phi}\equiv \frac{\partial W}{\partial \Phi}$.
We write the metric  as
\beqs
\di s^2_5&=&\di r^2+e^{2A}\di x_{1,3}^2\,,
\eeqs
and search for classical solutions of the form $A=A(r)$ and $\Phi=\bar{\Phi}(r)$,
 manifestly preserving Lorentz invariance.

Any solutions to the first-order equations
\beqs
A^{\prime}&=&-\frac{2}{3}W\,,~~~~~~~~\bar{\Phi}^{\,\prime}\,=\, W_{ \Phi}\,,
\eeqs
where $^{\prime}$ denotes derivatives in respect to $r$,
solve also the full set of coupled second-order differential equations of the five-dimensional system.
The solution of the first-order equations is 
\beqs
\bar{\Phi}(r)&=&\sqrt{\frac{3}{\Delta}}{\rm arctanh}\left(\frac{}{}e^{-\Delta(r-c_1)}\right)\,,\\
A(r)&=&A_0+\frac{1}{2\Delta}\ln\left(\frac{}{}-1+e^{2\Delta(r-c_1)}\right)
\ .
\eeqs
We will be interested in ratios of masses, and hence we set the two integration constants to
$c_1=0$ and $A_0=0$. In the UV (for large $r$) the background is
asymptotically AdS$_5$ with $A\simeq r$, and $\bar{\Phi}\sim
\sqrt{\frac{3}{\Delta}}e^{-\Delta r}$, while the space ends at $r=0$. 

From the UV expansion, one sees that  $m\equiv \sqrt{\frac{3}{\Delta}}e^{\Delta c_1}$ is the
dimensionful parameter that introduces a scale, analogous to the mass
deformation $m$ in Sec.~\ref{Sec:lattice}. 
This deformation makes the space end at $r\rightarrow c_1$,
and yields a mass gap in the spectrum, controlled by a scale $\Lambda_0$, in the dual theory.
Setting $m$ to zero ($c_1\rightarrow -\infty$)
would yield the exact AdS$_5$ background that is the gravity dual of a CFT. 
We are modelling a scenario that is qualitatively the same as in the lattice calculations in Sec.~\ref{Sec:lattice},
except for the fact that the gravity dual
cannot be used to describe a weakly-coupled fixed point, and hence  $\Lambda_{\ast}\rightarrow +\infty$.

We compute the spectrum of scalar and tensor fluctuations, that we interpret in terms of 
the glueballs of the (putative) dual field theory, by deriving the
linearised equations of motions around the background solution, and by
constructing explicitly gauge-invariant combinations of the
fluctuations~\cite{BHM, E}. 

We  introduce two cutoffs, by restricting the radial direction
to the segment $0<r_{\mathrm{IR}}<r<r_{\mathrm{UV}}$. 
$r_{\mathrm{IR}}$ plays the  role of $\Lambda_{\mathrm{IR}}$ (the lattice volume $L$),
while $r_{\mathrm{UV}}$ corresponds to $\Lambda_{\mathrm{UV}}$ (the lattice spacing $a$) in Sec.~\ref{Sec:lattice}.

Notice how the function $A(r)$ is monotonic and takes values from $-\infty$ for $r\rightarrow 0$ to $+\infty$ 
for $r\rightarrow +\infty$. We indicatively identify the cutoff scales as
\beqs
\frac{\Lambda_0}{\Lambda_{\mathrm{IR,UV}}}& \equiv & e^{-A(r_{\mathrm{IR,UV}})}\,.
\eeqs

After Fourier transforming,
the bulk equation
for the gauge-invariant scalar 
 $\mathfrak{a}(r,q_{\mu})\equiv\varphi-\frac{\bar{\Phi}^{\prime}}{6A^{\prime}} h$ 
is~\cite{E}
\beqs
\left[\frac{}{}\partial_r^2+4A^{\prime}\partial_r+e^{-2A}M^2\right]\mathfrak{a}
-\left[V_{\Phi\Phi}
+\frac{8\bar{\Phi}^{\,\prime}V_{\Phi}}{3A^{\prime}}
+\frac{16V\bar{\Phi}^{\prime\,2}}{9A^{\prime\,2}}\right]\mathfrak{a}&=&0\,.
\eeqs

We define $M^2\equiv -\eta^{\mu\nu}q_{\mu}q_{\nu}$,
in terms of the 4-momentum $q^{\mu}$.
We impose boundary conditions  according to~\cite{EP},\footnote{We take to $+\infty$ two boundary mass terms 
that are allowed by the symmetries of the model~\cite{EP}.
This procedure is equivalent, in the present context, to requiring regularity and normalisability, 
according to the standard prescription of gauge/gravity dualities.}
and  repeat the calculation  to extrapolate the results to the (physical) case $r_{\mathrm{IR}}\rightarrow 0$ and $r_{\mathrm{UV}}\rightarrow +\infty$.
The physical  results do not depend on the spurious regulators  $r_i$.
The boundary conditions are~\cite{EP} 
\beqs
\left.\left[\partial_r
+\frac{M^2}{e^{2A}}\frac{3A^{\prime}}{2\bar{\Phi}^{\prime\,2}}-\left(\frac{4V\bar{\Phi}^{\,\prime}}{3A^{\prime}}+V_{\Phi}\right)\right]
\mathfrak{a}\,\right|_{r_i}&=&0\,.
\eeqs

The traceless transverse components $\mathfrak{e}^{\mu}_{\,\,\,\nu}$ of the fluctuations of the metric 
obey the same equations as a scalar field with canonical kinetic term and no potential~\cite{MP}:
\beqs
\left[\frac{}{}\partial_r^2+4A^{\prime}\partial_r+e^{-2A}M^2\right]\mathfrak{e}^{\mu}_{\,\,\,\nu}&=&0\,.
\eeqs
We impose Neumann boundary conditions at the boundaries:
\beqs
\left.\partial_r\mathfrak{e}^{\mu}_{\,\,\,\nu}\,\right|_{r_i}&=&0\,.
\eeqs

\begin{figure}[t]
\includegraphics[width=0.95\columnwidth]{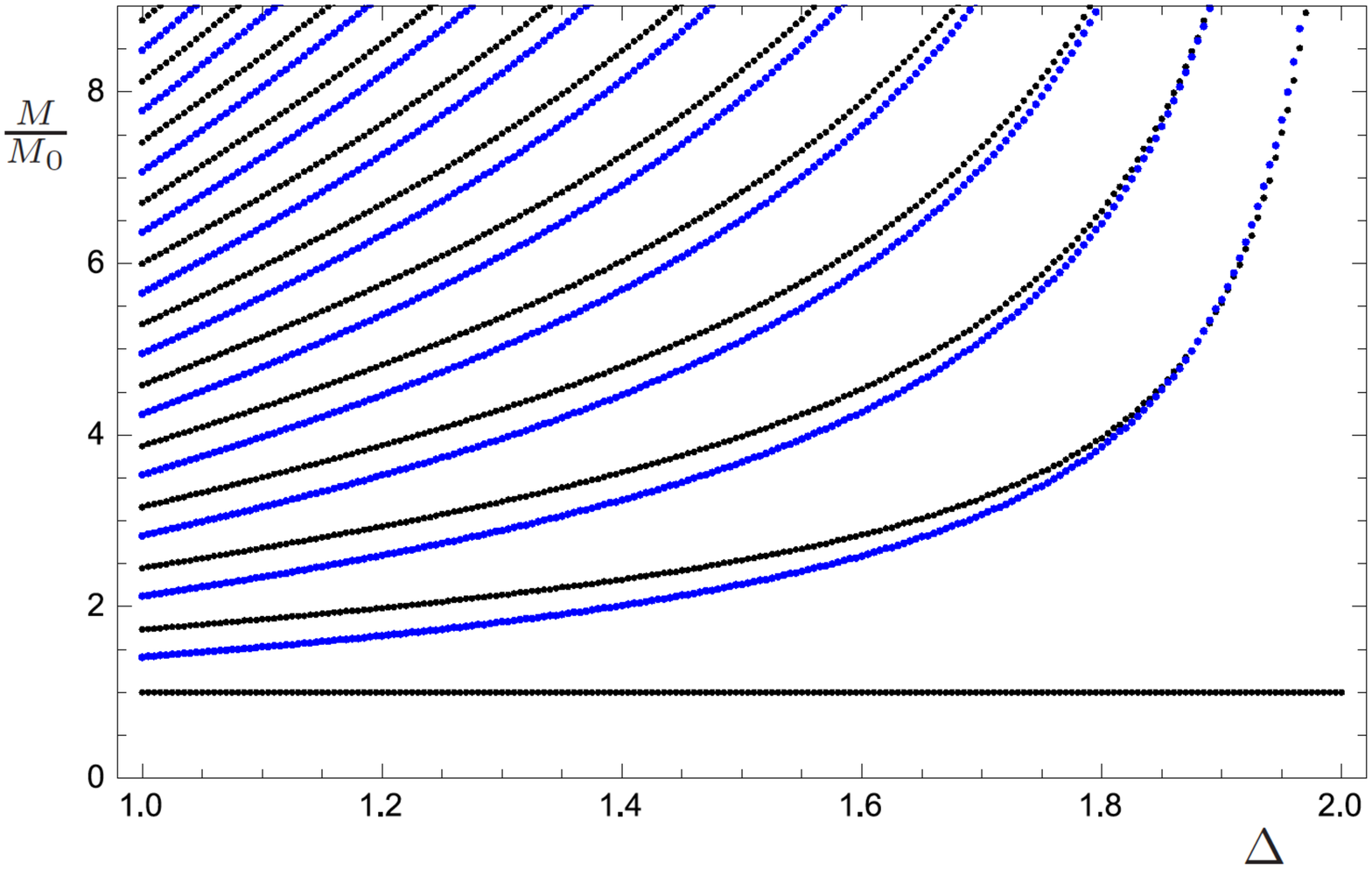}
\caption{The mass $M$ of the composite spin-0 (black) and spin-2 (blue) states, as well as their excitations, 
computed for $c_1=0=A_0$,
as a function of $\Delta$, obtained for  $r_{\mathrm{UV}}=25$ and $r_{\mathrm{IR}}=10^{-6}$,
normalised to the mass $M_0$ of the lightest scalar.
The lowest blue line is also the ratio $R$.
}
\label{Fig:mg}
\end{figure}

There is only one physical scale, fixed by the deformation itself --- equivalently, by the end of space $c_1=0$, or
by the scale $\Lambda_{0}$.
The spectrum of scalar and tensor modes is a function 
of the one  free parameter $\Delta$. We focus on the range $1<\Delta<2$.
The results are shown in Fig.~\ref{Fig:mg}.
In particular, for $\Delta=1$ we find $R\equiv M_T/M_0 = \sqrt{2}$,
which reproduces  the GPPZ case~\cite{MP,ACEP}.

From the numerical study, we find that  for $\Delta=1.371(20)$ we have $R\simeq1.95(4)$,
obtained with $r_{\mathrm{IR}}=10^{-6}$ and $r_{\mathrm{UV}}=20$.
For $\Delta=1.925(25)$ the calculation requires to use higher values of the UV cutoff,
because of the proximity to $\Delta\rightarrow 2$. We find $R\simeq 6.53^{+1.50}_{-0.91}$
with $r_{\mathrm{IR}}=10^{-6}$ and $r_{\mathrm{UV}}=50$.
These results are the pink shaded regions in Figs.~\ref{fig:fitrvsm0PP_2} and~\ref{fig:fitrvsm0PP_1},
which can be thought of as indicative predictions from the gravity calculations.

{
In proximity of $\Delta=2$,  the mass of the lightest scalar state
approaches zero. The $\Delta=2$ case is special because it corresponds to saturating the 
Breitenlohner-Freedman bound~\cite{BF}, in proximity of which 
non-trivial phenomena appear (see for instance~\cite{CL}).}

In addition, when $\Delta\simeq 2$ the mass of the tensor modes becomes approximately 
degenerate with the masses of the excited scalars. This is probably accidental,
yet it might be relevant phenomenologically. In some of the analysis in~\cite{diphoton}
the hypothesis that the diphoton resonance has a comparatively large width 
seems to improve the fit to the data.
If the diphoton signal were to be interpreted in terms of two new narrow resonances,
one with spin 2 and the other with spin 0, with masses close to one another,
the  large visible width  would be easier to explain.

\subsection{Cutoff effects}
\label{Sec:gravityartefacts}

We perform here an exercise aimed at illustrating 
some subtleties related to the role of the cutoffs used in the calculation of the spectrum.
This discussion is intended to be read in parallel with Sec.~\ref{Sec:lattice}.

There are three scales in the gravity calculation of the mass spectrum.
One is the physical scale $\Lambda_0$ induced non-trivially  by the deformation $m$. 
The other two are spurious scales,  due to the finite values of $r_{\mathrm{IR}}$ and $r_{\mathrm{UV}}$ --- corresponding to $\Lambda_{\mathrm{IR}}$ and 
$\Lambda_{\mathrm{UV}}$, respectively.
One must check that by repeating the calculations with larger and larger $r_{\mathrm{UV}}$ (smaller and smaller $r_{\mathrm{IR}}$),
eventually the results become insensitive to $r_{\mathrm{UV}}$ ($r_{\mathrm{IR}}$). 
Furthermore, this must be true while varying independently the 
three mass scales.
This is how  Fig.~\ref{Fig:mg} has been obtained.

\begin{figure}[t]
\includegraphics[width=0.95\columnwidth]{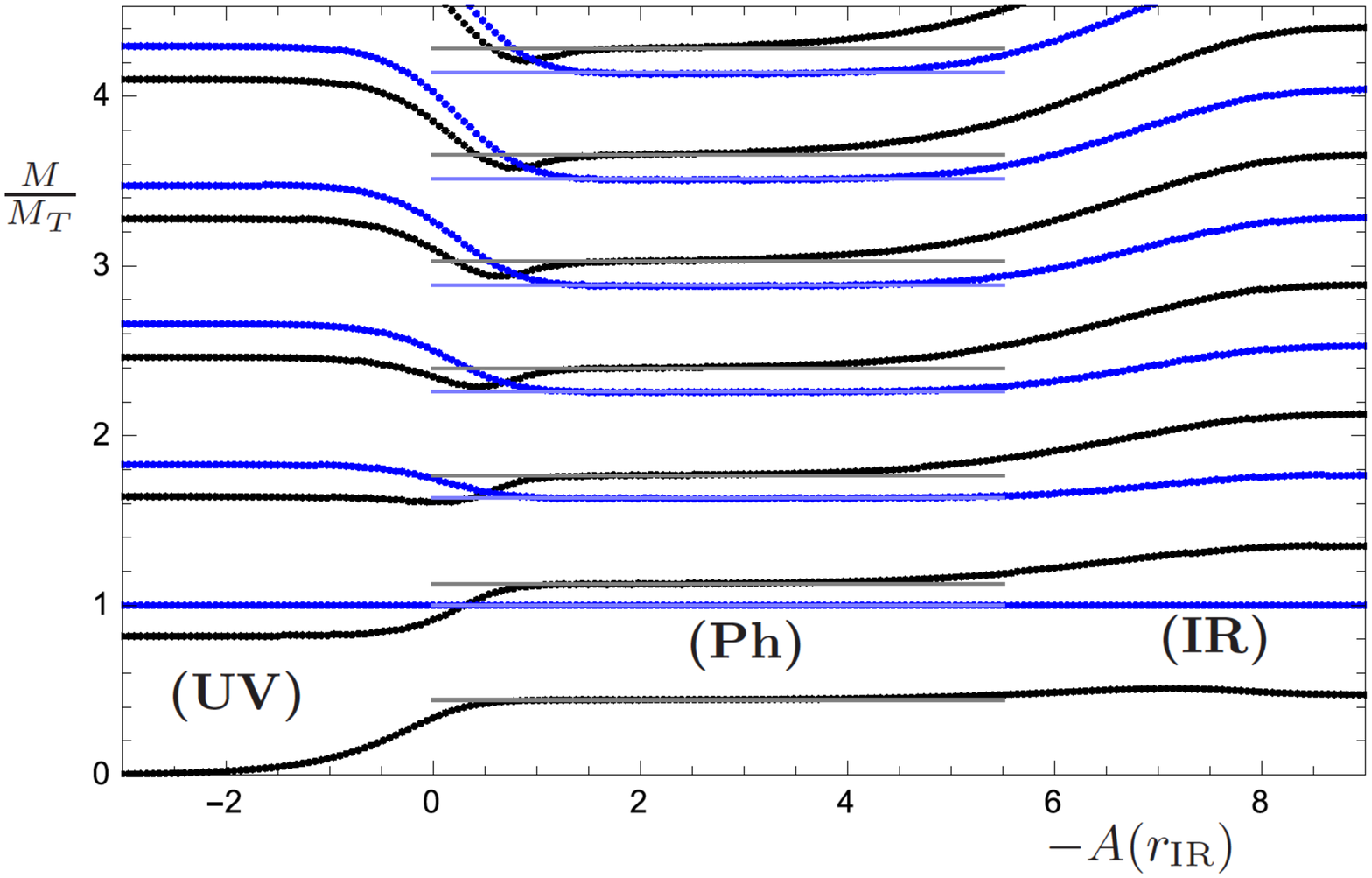}
\caption{ The mass $M$ of the composite spin-0 (black) and spin-2 (blue) states, computed for $c_1=0=A_0$,
as a function of $-A(r_{\mathrm{IR}})=\log (\Lambda_0/\Lambda_{\mathrm{IR}})$, for $\Delta=1.5$. We also vary the UV
cutoff, by keeping fixed  $A(r_{\mathrm{UV}})-A(r_{\mathrm{IR}})=8$.
 $M_T$ is the mass of the lightest tensor.
The straight lines  
are the result from Fig.~\ref{Fig:mg}, showing agreement between the two calculations in the (Ph) region,
as explained in the text.
}
\label{Fig:CUT8}
\end{figure}

Fig.~\ref{Fig:CUT8} illustrates the artificial distortions of the spectrum  in the presence 
 of correlated finite cutoffs.
We fix  $c_1=0=A_0$ and compute the spectra for $\Delta=1.5$, 
by varying both the IR and UV cutoffs,
but imposing the constraint that 
$A(r_{\mathrm{UV}})-A(r_{\mathrm{IR}})=8$.

The ratio $\Lambda_0/\Lambda_{\mathrm{IR}}$
 defines the IR cutoff.
The  spectrum is normalised to the mass $M_T$ of the lightest tensor excitation. 
The horizontal axis of the plots has the same meaning as in Figs.~\ref{fig:fitrvsm0PP_2} and~\ref{fig:fitrvsm0PP_1},
as $\Lambda_0/\Lambda_{\mathrm{IR}}\propto L M_0$.

There are in general three regions of $\Lambda_0/\Lambda_{\mathrm{IR}}$,
when the separation between the cutoffs is kept finite.
\begin{itemize}

\item[(UV)] For small $\Lambda_0/\Lambda_{\mathrm{IR}}$ the gravity calculations
probe only a region where the background is close to AdS.

\item[(IR)]  For large $\Lambda_0/\Lambda_{\mathrm{IR}}$, because the ratio of cutoff scales is kept fixed,
the calculation probes only the region close to the end-of-space,  far from AdS.

\item[(Ph)]  For intermediate values of $\Lambda_0/\Lambda_{\mathrm{IR}}$, and provided the ratio of cutoffs is large,
 the numerical study is probing both asymptotic regions of the geometry. 

\end{itemize}

The physical region (Ph) is the one corresponding to Fig.~\ref{Fig:mg}. If
 we take $\Lambda_{\mathrm{IR}}$ small enough and $\Lambda_{\mathrm{UV}}$ large enough,
the gravity calculation is sensitive to both the physics of the (dual) fixed point --- in particular to $\Delta$ --- as well as to confinement.
In Fig.~\ref{Fig:CUT8} this region is reached in the middle of the plot:
the ratio of the  lowest blue line and lowest black line agrees with $R\simeq 2.2$ from Fig.~\ref{Fig:mg} for $\Delta=1.5$.
In this region, universality is expected to play a role, and hence gravity, lattice and field theory to yield the same results.
The width of this intermediate plateau depends on how far separated the cutoffs are.
In particular, this plateau disappears if one takes the cutoffs too close to one another.

In the (IR) region the calculation is performed with cutoffs that are both near enough the end  of space
to be insensitive to the AdS region
--- see  the right part of Fig.~\ref{Fig:CUT8}.
The results are completely unphysical, affected by the artificial modelling of the mass gap.
The comparison to the right-end part of 
Figs.~\ref{fig:fitrvsm0PP_2} and~\ref{fig:fitrvsm0PP_1}
has to be done carefully.
In this region the lattice calculation agrees with the $\Delta=1$ gravity calculation, in which
 $R=\sqrt{2}$ (see the discussion in Section~\ref{Sec:YM}).
This is not the result shown in Fig.~\ref{Fig:CUT8}, which is performed with $\Delta=1.5$,
and entirely contained in a region of the geometry in which the gravity calculation cannot be trusted.
 
In the (UV) region the  spectrum results from a finite-volume artefact:
 an approximately scale-invariant theory is
forced inside a small box. The most striking feature 
 is the presence of a parametrically light scalar state.
Finite volume introduces  a  VEV,
spontaneously breaking scale invariance, and hence yields a Goldstone boson (dilaton).
In the limit of exact CFT, this state would be exactly massless, and decoupled from physical 
correlation functions. For finite $\Lambda_0/\Lambda_{\mathrm{IR}}$, the deformation  
provides both mass and  couplings to the dilaton. 

The lightness and small coupling of the first scalar imply that either 
one gets arbitrarily large $R=M_T/M_0$,
or  one computes $R$ from the second scalar excitation.
In the latter case, $R$ is smaller than in the (Ph) region, 
eventually approaching $R\simeq 1$, along the lines of the 
femto-universe~\cite{F,L}.
The specific details are model-dependent (not universal).

It is interesting to notice that some semi-quantitative cutoff features  in gravity reproduce those of 
 lattice calculations even in the (UV) region.
First of all, $R<\sqrt{2}$ (which is obtained ignoring the lightest
scalar) is close to $1$~\cite{FMPPRTZ}, as can be seen from Fig.~\ref{Fig:CUT8},
as well as in the region with $L M_0<8$ of Fig.~\ref{fig:fitrvsm0PP_2}.
Secondly, at least in some regions of parameter space,  $M_T<M_0$, as seen in
Fig.~\ref{Fig:CUT8} as well as in the perturbative calculation in~\cite{F}.
Thirdly, at large $\Lambda_{\mathrm{UV}}/\Lambda_{\mathrm{IR}}$ one finds $R\simeq 1.2$,
similar to the lattice strong-coupling expansion~\cite{Munster}.
It would require a dedicated study to establish  how many of these observations are 
more than just
accidental results.

\subsection{Comments about pure Yang-Mills theory}
\label{Sec:YM}

We notice the striking coincidence that three different theories
agree on   $R\simeq\sqrt{2}$: the $SU(2)$ Yang-Mills theory~\cite{Lucini:2001ej}, 
the $SU(2)$ theory with adjoint fermions~\cite{DelDebbio:2009fd,DelDebbio:2010hx}
in the regime in which $L M_0$ is large, and the  GPPZ  gravity model~\cite{GPPZ,MP}.
These independent results are not original to this work, yet the level of agreement is so good that it
deserves some further independent discussion. 
It is especially surprising that the agreement extends to the GPPZ model.
We devote this digressive subsection
to further suggest that  this might be the consequence of possible underlying universality
properties.

Among the many attempts to describe within gravity  a confining, large-$N$ Yang-Mills-like theory,
we would like to highlight three special ones. In  GPPZ~\cite{GPPZ,MP}, the asymptotic (in the UV)
geometry  is AdS$_5\times S^5$  and  $R=\sqrt{2}$.
A less known example exists~\cite{WY} (see also~\cite{EFHMP}),
in which the local geometry is asymptotically Ad$S_6\times S^4$ (one dimension is compactified on a shrinking circle), 
and for which $R\simeq 1.6$.
In the Witten model~\cite{W}, the local asymptotic geometry is
Ad$S_7\times S^4$ (two dimensions are compactified on shrinking circles), and $R\simeq 1.7$~\cite{BMT}. 
Ad$S_{d+1}$ gravity backgrounds provide the dual of $d$-dimensional 
CFTs, and hence the three examples we reported are related (in the far UV) to CFTs living in $4$, $5$ and $6$ dimensions, respectively.
Coincidentally, notice that in the three cases $R\simeq \sqrt{d/2}$.

Large-$N$ theories should not agree with $SU(2)$  in four dimensions.
But if the ratio $R$ is universal, this special observable should depend only on universal quantities,
such as the dimensionality of the space-time, or the dimensions of the relevant operators defined by the CFT,
and not on  microscopic details.

The  extrapolation of lattice $SU(N)$ data is compatible with the prediction from the Witten model.
Yet the comparison with GPPZ --- the dual of which is a four-dimensional CFT,
and in which the anomalous dimension $\gamma^{\ast}=0$ --- suggests that at least in the proximity of
the fixed point (which seems to be crucial in computing $R$, as our study shows)
 the dynamics of Yang-Mills theory is better captured by  GPPZ.
Even some field theory arguments in~\cite{B} --- in which the author also draws a comparison with large-$N$ studies,
but without relying on infinite-$N$ extrapolation --- yield $R=\sqrt{2}$. 

The remarkable coincidence  on $R=\sqrt{2}$
 between several different 
lattice, field theory and gravity calculations, at large-$N$ as well as at small-$N$,
for theories in which anomalous dimensions are trivial,
 might be the result of deep universality properties
of the ratio $R$, although a firm conclusion is premature.

\section{Physics Lessons}
\label{Sec:physics}

In this paper, we are looking for an example of a strongly-coupled field theory
yielding the ratio of masses $R\gg 1$. Besides intrinsic theoretical reasons,
we are interested to find evidence of this because
the recent signals of a new particle with mass $M\simeq 750 \gg 125$ GeV 
might admit an explanation in terms
of a heavy  glueball in a strongly-coupled extension of the Standard Model.

The working hypotheses under which we discuss  lattice data on $SU(2)$ with adjoint matter are the following.
\begin{itemize}

\item The $SU(2)$ theory with one or two adjoint matter fields is asymptotically free and close to conformal in the IR.

\item In it, the masses of composite states  shows a universal scaling with $m$, with
scaling exponent determined by the anomalous dimension of the $\bar{\psi}\psi$ operator.

\item While the coefficients in front of these  scaling laws are model-dependent, certain ratios, in particular $R$,
exhibit universal characteristics.

\end{itemize}

In this Section, we critically discuss whether the numerical studies 
support these hypotheses.
We want to disentangle physical results from lattice artefacts
and assess whether there are indications that $R\gg 1$.

We are aided in our task by the intuition gained in Section~\ref{Sec:string}.
The toy model in the gauge/gravity context suggests that the ratio $R$ is a monotonic function
of $\Delta=1+\gamma^{\ast}$, with a minimum  $R=\sqrt{2}$ for $\Delta=1$.
$R$ diverges at $\Delta\rightarrow 2$. This fact indicates that 
for gauge theories with large anomalous dimensions one should find $R\gg 1$.

We want to check whether the behaviour indicated by the toy model is confirmed
by the more rigorous and concrete lattice calculations. We start from the case of $N_f=2$
in Fig.~\ref{fig:fitrvsm0PP_2},  that can be summarised by three possible values of $R$.
For the largest values of $L M_0$ the results agree with the
Yang-Mills theory value $R = 1.44(4)$, 
as well as the gravity result for $\Delta=1$ --- in GPPZ, $R=\sqrt{2}$.
For intermediate values of $L M_0$, $R$ is close (within errors) to the prediction of the gravity toy model $R\simeq 1.95(4)$.
For smaller values of $L M_{0}$, $R$ is visibly lower, with the
decrease likely to be a sign of the approach to the femto-universe regime.

The large $L M_0$ region is interpreted in terms of artefacts that make the system lose memory
of the presence of the IR fixed point.
We interpret the intermediate one as the physical region, in which the lattice results can be extrapolated to the continuum limit.
The low $L M_0$ region is interpreted as a lattice artefact: given that the value of $\beta$ is fixed, and that the maximum number of lattice sites $n$
is finite, for small values of $L M_0$ we are exploring the region of (lattice) parameter space in which $m$ is so small that the 
discretisation of the spectrum is due to the finite volume effects, with $\Lambda_0\ll\Lambda_{\mathrm{IR}}$, 
 as in the femto-universe.

This interpretation is in line with what is shown in Fig.~\ref{Fig:CUT8}, 
obtained from gravity,
in spite of the fact that 
cutoff effects 
are not universal.
A non-trivial test of our interpretation would require a more detailed and precise measurement of the spectrum of scalars at 
small values of $m$, with higher statistics. 
Finite volume effects should be associated with the appearance of a spurious, weakly-coupled,  light spin-0 state, that
may have escaped detection. 

The study of  $SU(2)$  gauge theory with two adjoint fermions does not contradict the 
results from  gravity. 
In the physical region (Ph)
we do see numerical evidence of an enhancement of the ratio $R$ in
respect to pure Yang-Mills theory,
by an amount compatible with  gravity results, though modest ---
 the anomalous dimension $\gamma^{\ast}$ is small.

In the case of one Dirac adjoint fermion,  the anomalous dimension is large, and $\Delta=1.925(25)$ is  close to
its natural upper bound. The gravity calculation yields large $R= 6.53^{+1.50}_{-0.91}$.
In Fig.~\ref{fig:fitrvsm0PP_1},  for large $L M_0$ the ratio $R$ is small, broadly speaking compatible with
the result of pure Yang-Mills $R= 1.44(4)$ (within the large errors).
Going to smaller $L M_0$, $R$ is growing monotonically until a maximum of $R\simeq 5.4$.

We interpret this behaviour in the following terms. 
For large $L M_0$ the numerical results are dominated by lattice artefacts that
hide the effects of the IR fixed point.
Going to smaller $L M_0$, the calculations are approaching the physical plateau at large $R$, but do not quite reach it.
The physically relevant value of $R$ is realised in a region of (lattice) parameter space
to the left of the points available to this study.

In order to assess whether this interpretation is correct, it would be necessary to perform 
additional lattice simulations, for larger lattices (bigger $n$), and  smaller values of the mass $m$.
Such a study should show the appearance of a plateau at  large $R$.
The reader should exercise caution in using the quantitative comparison to the gravity toy model.
For $\Delta$ close to its natural maximum value $\Delta\sim 2$, the ratio $R$ is predicted to diverge.
This means that the result is very sensitive to the exact value of $\Delta$ itself, or equivalently of the anomalous dimension
$\gamma^{\ast}$, the precise determination of which is non-trivial.

There is an alternative logical possibility in interpreting Fig.~\ref{fig:fitrvsm0PP_1}, in which the raise in $R$
is attributed to a lattice artefact --- the large volume effects that introduce a spurious light scalar in the spectrum.
This is the same scalar we claimed to be hard to identify in the
$N_f=2$ case. For this reason, and because we do not see a plateau at a large value of $R$, we discard this
possibility.

In summary, the lattice results are consistent with our working hypothesis.
We find evidence of an enhancement of $R$ for the theory in which the
IR fixed point has large anomalous dimensions.
The enhancement we see is compatible with the prediction from gravity.
Accidentally, $R$ comes close to what would be needed to interpret the diphoton signal ($R\sim 6$).

We conclude with a cautionary remark about phenomenological applications.
The themes of this paper are to enquire on whether $R$ exhibits a
universal character in diverse theories and whether there exist theories for which $R\gg 1$.
We focus on $R$ because this quantity is well defined in a  broad 
class of field theories, and not directly affected by model-dependent details.
In particular, we exhibited explicitly the chiral symmetry and chiral-symmetry breaking pattern 
of the two $SU(2)$ theories, showing that they are very different. And we compared to 
a gravity model in which there is no chiral symmetry at all.
The fact that we find large values of $R$ for the $N_f=1$ theory is encouraging for phenomenological purposes,
in particular in reference to the diphoton anomaly at the LHC,
as a step towards a proof of principle that large mass hierarchies can arise in strongly-coupled theories.

Conversely, Fig.~5 of~\cite{Athenodorou:2014eua} shows that the spectrum of mesons
includes several particles that are lighter than the tensor glueball, and do not correspond to 
states observed at the LHC.
The phenomenological viability of any model requires studying carefully many other model-dependent 
details that go beyond our present purposes, including the task of finding a strongly-coupled model that
has large anomalous dimensions without introducing in the spectrum a plethora of light mesons.

\section{Conclusions and Outlook}
\label{Sec:outlook}

We have exhibited what is, to the best of our knowledge, the first example of
a lattice study of
a strongly-coupled theory in which the ratio of  tensor to scalar glueball mass
is large ($R>5$). This is  encouraging, in the light of  the recently observed
LHC diphoton excess, which would require $R\sim 6$.

The model in which these indications arise is the mass deformation of
a gauge theory with $SU(2)$ gauge group and $N_f=1$ generations
of Dirac fermions in the adjoint representation. In the massless limit, 
this theory is believed to have an IR fixed point with  large anomalous dimensions.
As a consistency check, we compared the numerical results to the lattice results for  the case $N_f=2$, in which the IR fixed point has small anomalous dimensions.
We found $R\sim {\cal O}(1)$, which is compatible with expectations.

We also  compared to a toy model built in the context of gauge/gravity dualities.
We drew a parallel between the cutoff effects on the lattice and in gravity models,
finding qualitative agreement.
In the physically relevant region of parameter space, we found
remarkably good numerical agreement with the lattice for the relation between $R$ and the dimension $\Delta$
of the deforming parameter.

These  are preliminary results, and require further investigation with dedicated  studies.
We summarise  some of the directions this research program could encompass.

On the lattice side, a dedicated program for the study of glueball spectra, both in $SU(2)$ with adjoint matter,
as well as in other candidate theories with IR conformal dynamics, is needed. 
Particular attention should be devoted to the $SU(2)$ theory with $N_f=1$, the only known case to date
for which $R$ is found to be large.  
New studies with larger lattices might
ascertain whether large values of $R$ are truly physical or affected by lattice artefacts.

On the gravity side, a more general exploration of models that describe the dynamics of deformations
of IR-conformal theories is needed. The results presented here are based on a simplified 
model built within the bottom-up approach to holography. 
It would be  useful to identify and study a full model derived from
string theory in the top-down context. Yet, it should be possible to
carry out similar studies  in more general contexts, within the
bottom-up approach.  

In the field theory context, it would be interesting to  study  the ratios of 
masses, aided by statistical field theory arguments. The claim of universality 
that underpins the comparison we make of lattice and gravity results
might be taken literally (in the sense that $R$ is only a function of $\Delta$), 
or just in a broad sense:  the behaviours 
emerging in different models  share qualitative features.
In this paper, we took the latter view, in the absence of a rigorous proof of a more robust relation,
though  our numerical results show such a good level of agreement with
gravity calculations that it might be an  indication of a more fundamental physical principle.

On the phenomenological side, several additional questions arise.
Under the assumption that  new strong dynamics can explain a large mass hierarchy  $R\gsim 5-6$, 
the construction of a realistic model of electroweak symmetry breaking
requires to introduce many  additional ingredients.
It is worth bringing to the attention of the reader
for example the papers in~\cite{VS}, where some relevant  considerations 
pertaining to model-building and LHC phenomenology are discussed.
This paper is a first step towards a future proof of principle that comparatively heavy
spin-2 resonances can emerge as composite states of new strong dynamics, leaving 
all  model-building and phenomenological issues aside.

{
We close by repeating the main results of this study.
We collected a significant body of  empirical evidence,
both from lattice studies of $SU(2)$ theories with adjoint matter and from simplified 
gravity dual models, that points in two intertwined directions.
First of all, we find agreement (within errors)  in the results for $R$ computed in completely different theories
that share the same dimension  $\Delta$ for the relevant deformation,
suggesting that the quantity $R$ might be a manifestation of some form of universality.
Secondly, when the deforming operator has large anomalous dimension,
we find a parametric enhancement of $R$, 
the ratio of masses of the lightest tensor and scalar glueballs.
Both results need to be further tested with more extensive, specific future studies.
}
\vspace{1.0cm}
\begin{acknowledgments}

We thank L.~Del Debbio, A.~Patella, A.~Rago and L. Vecchi for useful comments on 
early versions of the manuscript, and J.-W.~Lee for useful discussions.

AA has been partially supported by an internal program of the
University of Cyprus under the name of BARYONS. In addition, AA
acknowledges the hospitality of the Cyprus Institute, where part of
this work was carried out.   

The work of CJDL is supported in part by the Taiwanese MoST grant
number 102-2112-M-009-002-MY3, as well as  by the OCEVU Labex
(ANR-11-LABX-0060) and the A*MIDEX project (ANR-11-IDEX-0001-02),
which are funded by the "Investissements d'Avenir" French government
programme and managed by the "Agence nationale de la recherche (ANR)''.

The work of DE is supported in part by 
 the National Research Foundation of South Africa (unique grant
number 93440). DE would also like to thank the hospitality of 
the University of Barcelona, where part of this work was completed.

Numerical computations were executed in part on the HPC
Wales systems, supported by the ERDF through the WEFO (part
of the Welsh Government), on the Blue Gene Q system at the Hartree
Centre (supported by STFC) and on the DiRAC Blue Gene Q
Shared Petaflop system at the University of Edinburgh. The latter is
operated by the Edinburgh Parallel Computing Centre on behalf of the STFC DiRAC HPC Facility
(www.dirac.ac.uk). This equipment was funded by BIS
National E-infrastructure capital grant ST/K000411/1,
STFC capital grant ST/H008845/1, and STFC DiRAC
Operations grants ST/K005804/1 and ST/K005790/1.
DiRAC is part of the National E-Infrastructure.  

The work of BL and MP is supported in part by WIMCS and by the STFC grant ST/L000369/1.

\end{acknowledgments}

\end{document}